\documentclass{nature}
\makeatletter
\usepackage{graphicx}
\let\saved@includegraphics\includegraphics
\AtBeginDocument{\let\includegraphics\saved@includegraphics}
\renewenvironment*{figure}{\@float{figure}}{\end@float}
\makeatother
\usepackage{caption}
\usepackage{pgfplots}

\title{Structural evolution and on-demand growth of artificial synapses via field-directed polymerization}
\author
{Matteo Cucchi,$^1$ Hans Kleemann,$^{1}$ Hsin Tseng,$^{1}$ Alexander Lee$^{1}$ \& Karl Leo$^1$}
\begin{document}
\maketitle
\begin{affiliations}
 \item Dresden Integrated Center for Applied Physics and Photonic Materials (IAPP), Technische Universit{\"a}t Dresden, 01062 Dresden, Germany
\end{affiliations}

\textbf{Interconnectivity, fault tolerance and dynamic evolution of the circuitry are long sought-after objectives of bio-inspired engineering. Here, we propose dendritic transistors composed of organic semiconductors as building blocks for neuromorphic computing. These devices, owning to their voltage-triggered growth and resemblance to neural structures, respond to action potentials to achieve complex brain-like features, such as Pavlovian learning, pattern recognition, and spike-timing-dependent plasticity. The dynamic formation of the connections is reminiscent of a biological learning mechanisms known as synaptogenesis, and it is carried out by an electrochemical reaction that we name field-directed polymerisation. We employ it to dendritic connections and, by modulating the growth parameters, control material properties such as the resistance and the time constants relevant for plasticity. We believe these results will inspire further research towards complex integration of polymerized synapses for brain-inspired computing.}
\begin{figure}
	\includegraphics[width=\linewidth]{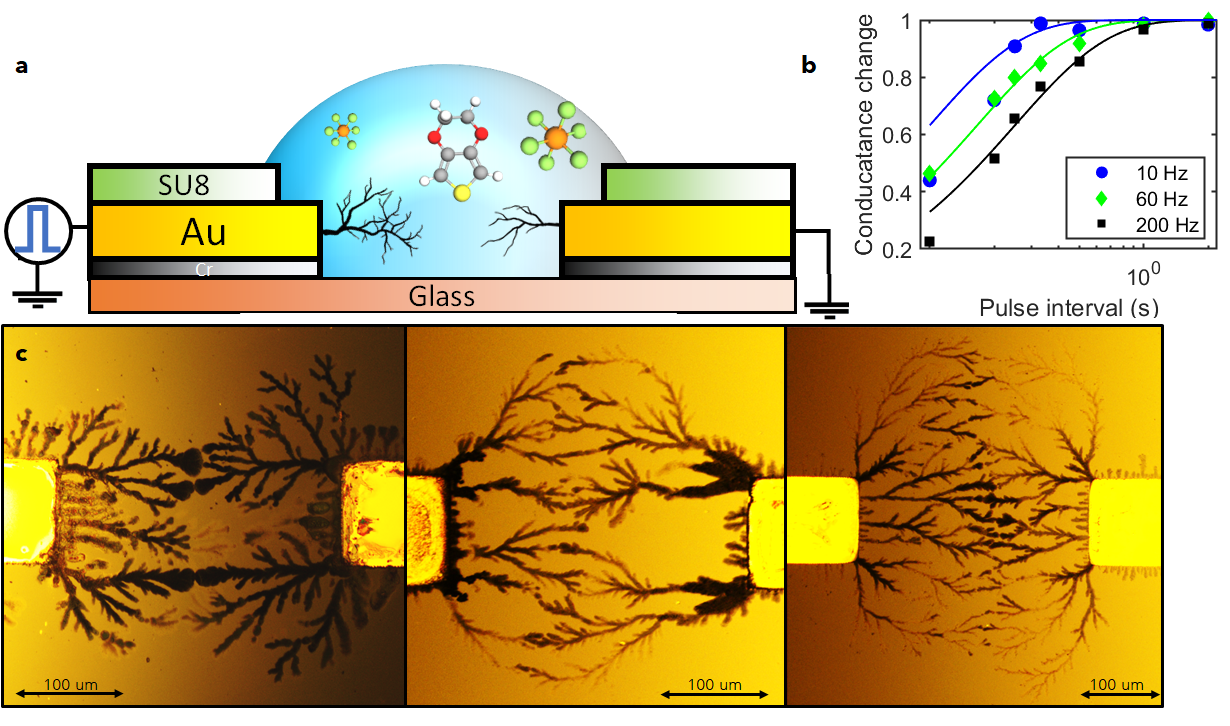}
	\caption{\textbf{Set-up and growth of the dendritic networks. a,} Setup used to grow the networks (for details see methods). \textbf{b,} By varying the polymerization frequency it's possible to control the capacitance of the channel. \textbf{c,} neural networks grown via FDP at 4 V with varying frequency: higher frequencies promote thinner fibres and a higher degree of branching.}
	\label{setup}
\end{figure}

The dense network of synapses and neurons constituting the human brain works in a fundamentally different manner from digital computers: while transistor-based devices excel in number crunching and binary logic, massive parallelisation  and continuous rewiring grant the brain superior pattern recognition and cognitive learning\cite{grossberg2000complementary}. The basic units of signal processing in the nervous system are the synapses: synapses electrically or chemically connect two neurons and, depending on the temporal pattern with which the pre- and post-synaptic neurons fire, they strengthen or weaken. It is then clear that sophisticated mechanisms control the outgrowth and branching of the synapses time-evolution. Such mechanism seems to be mediated by the formation of endogenous electric fields \cite{mccaig2000neurotrophins,mccaig1997physiological,patel1984perturbation}. On this basis, developing techniques that address the emulation of these brain-like process becomes attractive. In general, the replication of brain functions on artificial devices, known as neuromorphic computing, has recently sparked significant interest\cite{davies2019benchmarks,van2018organic}. 
Neuromorphic devices such as memristors\cite{strukov2008missing,xia2019memristive}, phase-change memories\cite{au2017phase} and floating-gate transistors \cite{danial2019two} were designed to show multiple ON states, potentiation, depression and neuroplasticity\cite{gkoupidenis2015synaptic,gkoupidenis2015neuromorphic}. These devices, inspired by the change of the number of neurotransmitters in chemical synapses, can achieve reversible modulation of their conductance.\\
Recently, devices based on organic semiconductors have drawn attention: their inherent softness and tunability, combined with the modest requirements in terms of electrical performance, make organic materials promising candidates for neuromorphic devices. Among these, organic electrochemical transistors show high biological resemblance as they work in electrolytic solutions\cite{khodagholy2013high}, analogous to humans' synapses, and with little power consumption. Moreover, bioinspired organic memristive devices and networks showed high potential for neuromorphic building blocks and impressive power efficiency\cite{erokhin2005hybrid,erokhin2010bio,van2017non}. Interestingly, Gerasimov $et$ $al.$, mimicked the long-term plasticity with OECTs by producing channels in response to electrical pulses \cite{gerasimov2019evolvable}. \\
However, no consideration was made about the exploitation of the branching and the directionality of the reaction to achieve complex neuromorphic functions. In general, mimicking the time-dependent evolution of the neuronal circuitry was never attempted. This dynamic rewiring, largely recognized as key to confer the brain its remarkable computational power, combined with high efficiency remains elusive on hardware \cite{cattell2012challenges}.\\
Here, we attempt to emulate the guided formation of synapses by using a technique we name field-directed polymerisation (FDP): by applying action potentials between two electrodes in a solution containing a monomer, we trigger the polymerisation of a conductive polymer that, following the field-lines, bridges the electrodes (see section 2 for detailed explanations). Each connection represents an artificial synapse. 
The polymer growth mimics the synaptogenic process, i.e. the biological process that underlies the formation of new synapses during the act of learning\cite{jan2010branching,jasinska2010rapid,kleim1996synaptogenesis,draganski2006temporal,sheng2018experience} (see Video S1) and the resulting networks resemble the dendritic topology of biological neuronal networks, while being fully functional as p-type OECTs (See Fig. S1).
We make use of the directionality of the reaction to grow synapses between two selected electrode (See Fig. S2); afterwards, once the channel is formed, it can be further reinforced via application of an AC field. We utilise the frequency modulation to control the conductance of the networks through structural modification, and to update the individual synaptic weight by controlling the degree of synaptic reinforcement. In this way, multiple ON states are achieved based on the learning time. Moreover, by varying the material morphology, the time constant that controls the plastic short-term effects of the channel is directly addressed.

  \begin{figure}
 	\centering
 	\includegraphics[width=\linewidth]{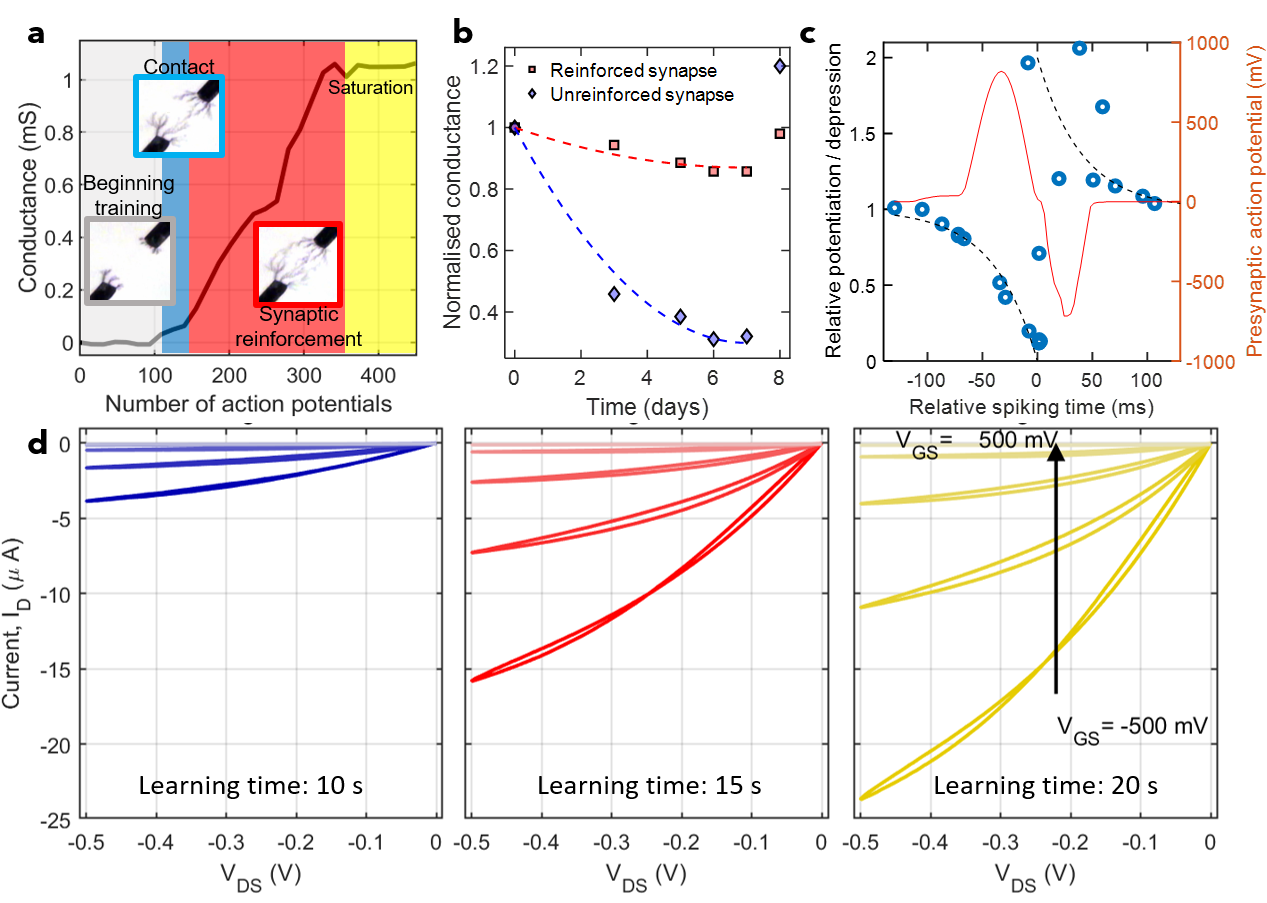}
 	\caption{\textbf{Structural modifications lead to learning and forgetting. a,} Learning curve during the formation of artificial synapses i.e. training. Blue area: prior to connection; red area: sharp increase of current indicating gap bridging; yellow area: more synapses/branches are formed. Inset figures taken from video S1 \textbf{b,} Normalised conductivity of a network grown to form different amount of synapses: the reinforced synapse shows a higher conductance, that slightly decreases after a week; unreinforced synapses show a steep decay. The last data point (day 8) is measured after newly reinforcing the connection. \textbf{c,} STDP behaviour of FDP-grown synapses. \textbf{d,} I-V characteristics of a single device measured at different training times (more details in SI).}
 	\label{fig:regrow}
 \end{figure}

\section{Field-directed polymerisation: growth mechanism}
Field-directed polymerisation (FDP) allows for the growth of polymeric fibres with a dendritic morphology. The synapse formation is carried out with the area between two electrodes immersed in a solution of acetonitrile containing 50 mM 3,4-Ethylenedioxythiophene (EDOT) and 1 mM tetrabutylammonium hexafluorophosphate (TBAPF). This recipe was suggested in ref. \cite{koizumi2016electropolymerization} in which a reducing agent is also added, that we found not strictly necessary. In addition, we employ much lower voltages (2 to 5 V).
Example of voltage-triggered polymerization are reported in literature, but rarely with an alternating signal \cite{eickenscheidt2019pulsed,gerasimov2019evolvable}, and never, to our knowledge, the material properties, branching degree and directed growth are used to build evolvable electronics or neuromorphic devices. 
The polymerisation is triggered by an AC bias applied across two arbitrary electrodes in the solution  (see Fig. \ref{setup}a). The AC amplitude applied must be sufficient to sustain the radicalisation of the monomer (EDOT) that is stabilised by the dopant (PF$_6^-$). We suggest that the reaction is a nucleation-and-precipitation process: during the positive part of the periodic waveform, radicals form at the interface with the electrode, where they drift away under the field or get neutralized by a dopant anion PF$_6^-$. In the latter case, oligomerization quickly occurs\cite{qiu1992electrochemically} and long, insoluble chains deposit at the interface (see Video S1). Fig. S3 shows a microscopic image of a fibre, with visible nucleation clusters.
 The time available for the reaction to occur is inversely proportional to the frequency of the signal, resulting in thin branches for high frequency (Fig. \ref{setup}c).
Also important is the waveform: to obtain dendritic shapes, the waveform shall show both positive and negative polarity, to attract and repel the ions sequentially. Interestingly, we can grow synapses using an amplified action potential waveform as shown in Fig. S4. These promising results might inspire research and development into new materials that can oxidatively polymerise at reduced voltages to interact with neural signals, grow in response to neuronal spikes and achieve neural interfacing.\\
The reaction always happens at the fiber extremity where, occasionally, bifurcation occurs. This stems from the higher local field (tip effect) that in turn accumulates more dopants. On the contrary, the field between two already-grown fibers is negligible (Faraday cage) and, even though monomers oxidise in this region, they drift away because no net dopant excess concentration exists.
\section{Long-term and short-term plasticity}
 The dynamic plasticity of biological synaptic networks occurs over many orders of magnitude in time - an essential property to execute complex tasks such as perception, computation, filtering, or long memory storage. The uniqueness of FDP allows for the formation of organic semiconductor-based artificial neural networks that can mimic dynamic plasticity over more than 9 orders of magnitude in time, from spike-timing-dependent plasticity in $\mu$s regime, over synaptic reinforcement at the second scale, to long term depression occurring at the scale of weeks (Fig. S5).
In Fig. \ref{fig:regrow}a, learning-induced synaptogenesis is mimicked: the conductance of the connection is enhanced by stimulation via action potentials. In the initial phase, the fibres have not yet bridged the gap between the electrodes and the measured current comes from the electrolytic solution. A sharp increase indicates the first connection.
Koizumi \textit{et al.} reported an abrupt termination of the growth upon contact\cite{koizumi2016electropolymerization}; however, we observe a prolonged branch formation and utilise this effect to further increase synapse conductance and emulate synaptic reinforcement, a key biological mechanism leading to memory encoding\cite{bailey2015structural} and long-term memory retention. The final saturation is achieved when the transport in solution becomes negligible compared to the current carried by the connecting fibers. The resulting S-shaped curve has potential applications in artificial neural networks, where self-limiting learning (saturation after synaptic reinforcement) is key to avoid overtraining\cite{tetko1995neural}, and the initial plateau ensures the presence of an activation threshold.  The fine-tuning of the electrical conductance through a structural/physical change of the fibres is further proved in Fig. \ref{fig:regrow}d. Here, the OECT characterization of  a dendritic connection after three sequential updates of its synaptic weight is reported.\\
As we learn new skills, we forget old ones: for the brain, forgetting is as essential as learning. Hence, it is mandatory to successfully implement a controlled forgetting/depression mechanism on brain-inspired hardware. A major process underlying the gradual erasure of long-term memories is synaptic decay\cite{hardt2013decay,sadeh2014we}. 
Here, we tune the decay time of a synapse by controlling the degree of synaptic reinforcement. In Fig. \ref{fig:regrow}b we grow and characterise the decay of two synapses with different degrees of synaptic reinforcement. A highly reinforced synapse with a dense fibre network ensures fault tolerance, increased conductance, and longer retention. In contrast, a weak connection witnesses a conductivity drop of 50\% in 48 hours. We assign the decay to the mechanical stress following the swelling of the PEDOT\cite{biessmann2018monitoring}: this causes the formation of large dopant domains with reduced doping efficiency.
On shorter timescales (milliseconds to minutes\cite{zucker2002short}), short-term plasticity assists the brain in computational tasks and ensures quick adaptation.
To achieve short-term depression, synaptic depression at the millisecond scale is emulated by employing a third electrode as a pre-synaptic neuron, with the source being the post-synaptic neuron. A pulsed signal  is applied, causing cations in the solution to accumulate within the synapse (Fig. S6) and dedoping the semiconducting polymer, thus modulating the source current. This effect, known as short-term depression (STD) is stronger if the pulses are applied in rapid sequence (due to an effect called pulsed-pair ? \cite{van2017non}, or weak if the ions have enough time to redistribute in the solution, causing the recovery of the original channel conductivity. Here, we vary the channel properties and morphology in order to control the time constant that governs STD. In Fig. 1b, STD is quantitatively reported as a function of the polymerisation frequency. For higher frequencies, the channel is easily depressed (halved for pulses separated by 200 ms). On the other hand, lower frequencies allow the formation of networks that, in order to undergo depression, must be subject to frequent pulses (for pulses separated by 200 ms its conductance decreases by 10\%). Such modulation of the post-synaptic neuron's conductance based on the temporal pattern of the presynaptic firing is analogous to biological nerves, and we highlight the impact of achieving control over the neuromorphic timescales via material parameters, that can be helpful in growing synapses and networks with specific time-dependent learning rules.\\
As an essential element of the short-term plasticity, we analyse the potentiation and depression of a synapse based on the relative spiking interval between the pre- and post-synaptic neuron. Such mechanism, referred in neuroscience as spike-timing-dependent plasticity (STDP), is at the heart of the potentiation/depression processes of a synapse. In simple words, if a presynaptic spike is likely to cause the post-synaptic neuron to fire, that synapse is going to reinforce; on the contrary, the synapse remains unaltered or weakens when such causality does not apply\cite{dan2004spike}. In OECTs, this mechanism is generally emulated by applying an action potential to the presynaptic neuron (gate), and another one of same amplitude but displaced in time, to the postsynaptic neuron (drain) \cite{alibart2012memristive,fu2018flexible}. The presynaptic signal modifies the ion distribution within the channel and increases or decreases the conductivity depending on the temporal offset. Fig. \ref{fig:regrow}c reports the modulation of the post-synaptic signal analogous to biological synapses, known as STDP.

\begin{figure}
	\includegraphics[width=\linewidth]{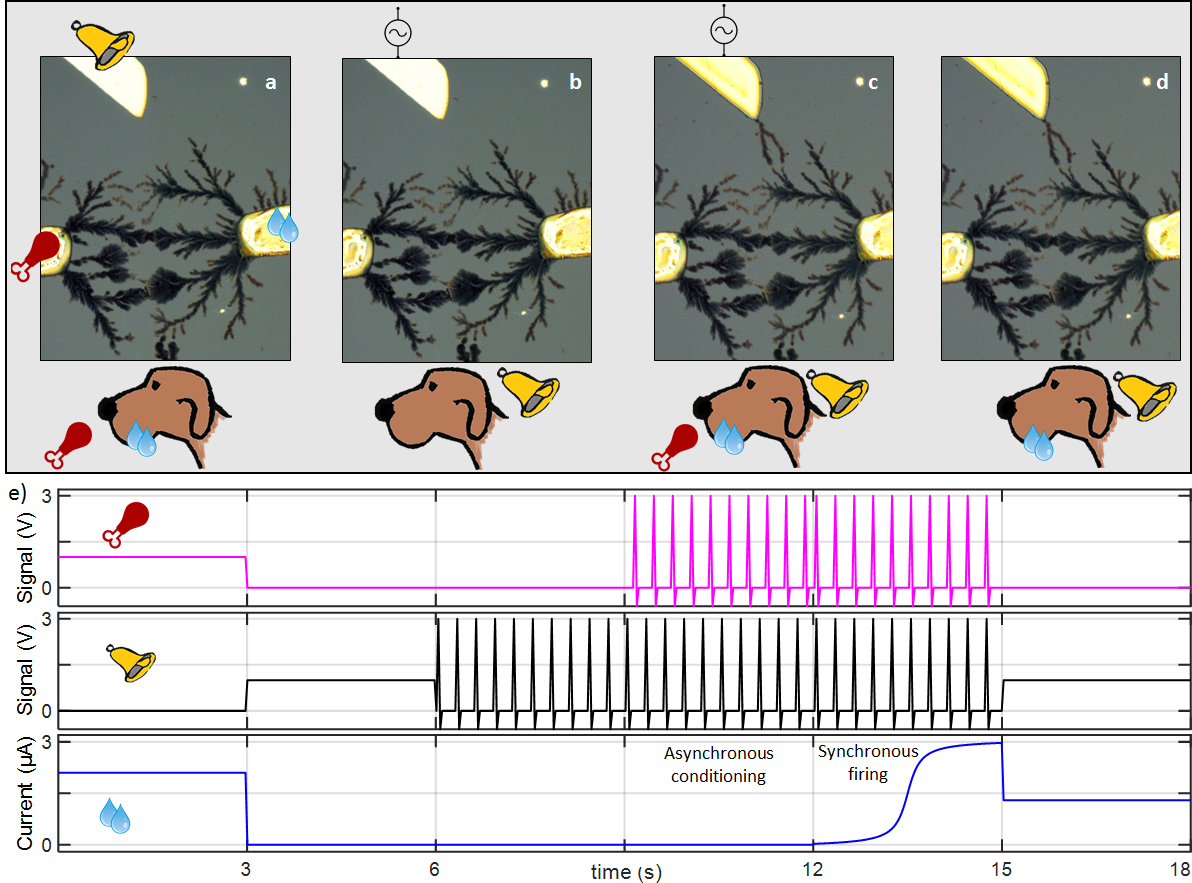}
		\caption{\textbf{Classical conditioning: the example of Pavlov's dog: a},t he electrode $food$ is connected electrically to the electrode $salivation$ while a third electrode $bell$ is not uncoupled initially. \textbf{b}, If the action potentials are applied to $bell$, the conditioning is unsuccessful. \textbf{c}, Only when the signal is applied to both the $bell$ and $food$ with the right temporal pattern, the learning mechanism begins and \textbf{d} results in the electrical contact between $bell$ and $salivation$. \textbf{e}, Key for a realistic emulation of the biological process is the period between 6 and 15 s: from 6 to 9s, the synapse does not grow when the pulse is applied only to the $bell$; from 9 to 12s, action potentials are applied both to $bell$ and $food$ with a time offset of 10 ms, resulting again in an unsuccessful training. The dog actually couple the $bell$ to the $salivation$ only when is exposed to $bell$ and $food$ simultaneously, as it happens between  12 and 15 s, in which the new synapse is generated.}
	\label{pavlov}
\end{figure}

\section{Pavlovian conditioning}
More complex neuromorphic features are attained by multiple electrodes. In this way, interconnectivity is implemented and the electric field in the solution is shaped to better direct the growth: this is achieved by using multiple electrodes at a potential lower than V$_{ox}$. In this way, the growth between two selected electrodes is allowed or prevented by the potential applied to neighboring electrodes (see scheme in Fig. S8). Furthermore, the need of such a threshold potential assures the presence of an activation function to begin the reaction; similarly to biological nerves, where the weighted sum of the inputs is passed through an activation function, here the summed effect of multiple electrode can trigger or inhibit the learning/growth process at one electrode.

We demonstrate the potential of such a feature by achieving classical conditioning: in 1903, Pavlov showed that the brain can pair a potent, hardwired, biological stimulus to initially neutral stimulus in a learning process that leads to the same physiological response when the subject is exposed to either stimulus\cite{pavlov1926conditioned}. To begin, an already existing connection conducts current between the "$food$" node and  the "$salivation$" node, while the "$bell$" node is initially decoupled from the system (Fig. \ref{pavlov}a): before conditioning, the dog instinctively salivates to the smell of food but does not respond to the ring of the bell and a signal to "$bell$" does not produce any output at "$salivation$" (0 to 3 s in Fig. \ref{pavlov}e).
To pair the two stimuli, a connection between "$bell$" and "$salivation$" must be produced. The bell ring is here represented as an action potential (2 V, 50 Hz); this signal alone does not suffice for polymerisation as the voltage is below V$_{ox}$ (Fig. \ref{pavlov}b). We thus have an initial independence of the two inputs, i.e. no conditioning occurs if the dog is exposed to the bell ring without food. The conditioning is also unsuccessful when both "$food$" and "$bell$" fire, but with a time offset of 10 ms. These two scenarios in which no potentiation occur are an essential feature in Pavlovian conditioning because they assure the necessity of a precise temporal synchronization for the synapse to undergo potentiation.
It is indeed successful when the firing is applied simultaneously i.e. the bell is rung when the food is served. This happens in panel c, where action potentials are applied between $bell$ and $salivation$, and between $food$ and $salivation$, causing the growth from $bell$ to $salivation$ (12-15 s). As a consequence of the training, a signal applied to either $bell$ or $food$ causes an output at $salivation$ (Fig. \ref{pavlov}d,e 15 to 18 s). The time-dependence of the conditioning is coherent with the Hebbian learning rule: "cells that fire together wire together".
The method described can be extended to a larger number of electrodes without requiring the complex circuitry needed for systems previously described\cite{ziegler2012electronic,van2017non}.  Notably, the lateral connection of synapses, such as the one in Fig. \ref{pavlov}d, emulates another well-know biological process, but poorly implemented on devices, known as heterosynaptic plasticity\cite{humeau2003presynaptic}.
\begin{figure}
	\centering
	\includegraphics[width=\linewidth]{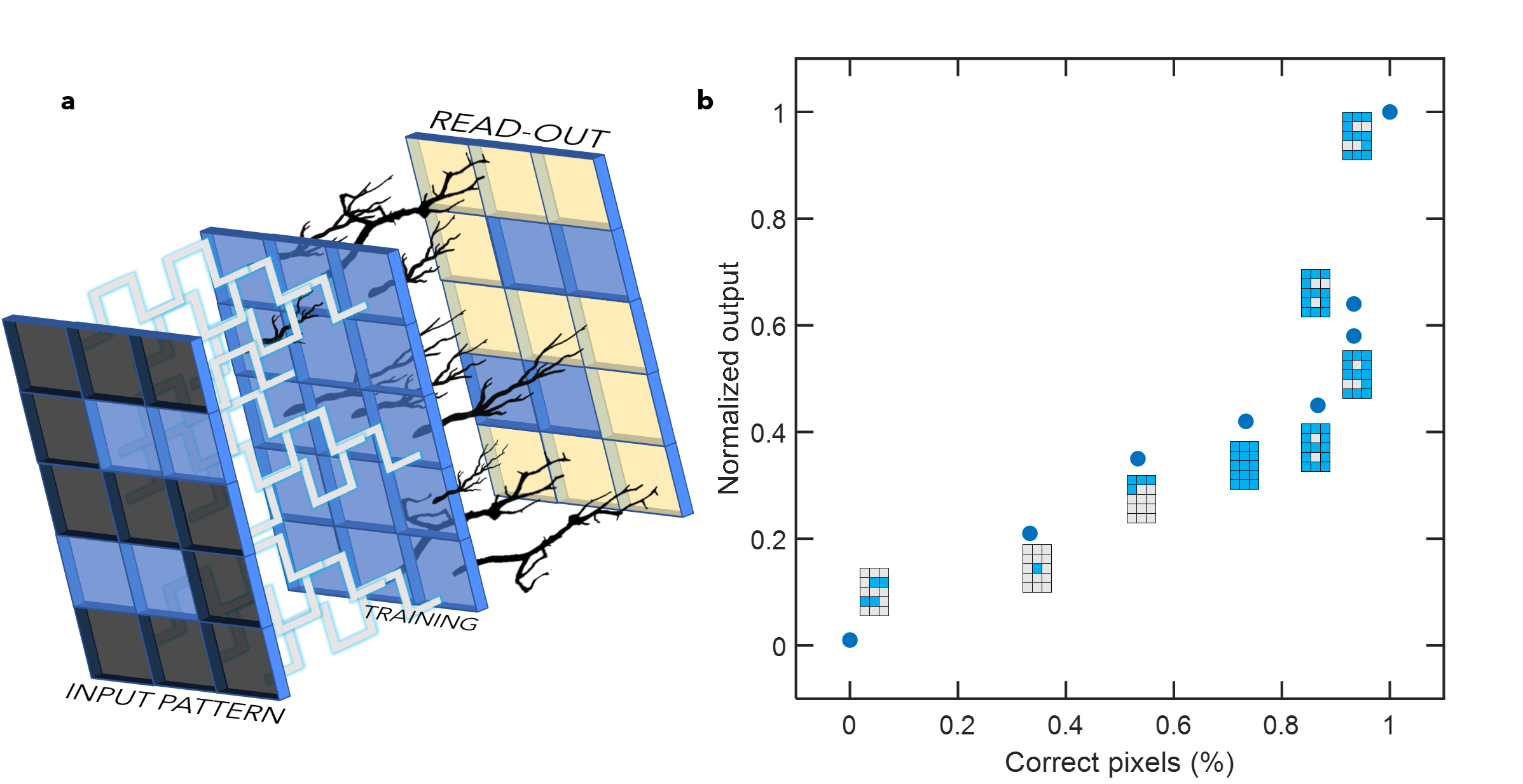}
	\caption{\textbf{Pattern recognition: a,} training: a signal corresponding to the pixel value is applied, causing the growth of the synapses where signal is applied. \textbf{b,} readout: The intensity of the signal correlates to the similarity of the trained input.}
	\label{fig:numberrecognition}
\end{figure}

\section{Pattern recognition}
We show the potential of FDP for in-memory neuromorphic computing by demonstrating a new device concept to train and recognise patterns. Here, we use a 3x5 pixel image as an input (the device can be extended to an arbitrary number of inputs/outputs). 
In the training phase, an action potential signal (50 Hz) is applied for 3 seconds to 15 input electrodes. The amplitude depends on the pixel value: 0 V if the pixel is void, and 3 V if the pixel is filled, as sketched in Fig. \ref{fig:numberrecognition}a, for the number 5. The voltage triggers the reaction and polymeric fibres grow towards the output electrodes (See Fig. S9 and methods for a detailed explanation).\\
The readout consists of the application of a constant bias or pulse to an arbitrary set of inputs while an aqueous solution of 1 mM NaCl immerses the network. The output current, corresponding to the recognition confidence, is integrated at the output nodes, and reported normalised in Fig. \ref{fig:numberrecognition}b.
The largest output is recorded when the input is equal to the trained pattern. One could add/remove pixels, thus reading out a wrong or semi-wrong pattern with respect to the one that was trained. Indeed, single-pixel error patterns (6 or 9 for instance) result in lower confidence while dissimilar inputs output even smaller values. This new device concept stems from the unique ability of FDP to grow networks composed of conductive/excitatory and inhibiting synapses: if a connection grows, input and output will be in electrical contact; if no connection is present at a specific input pad after the learning epoch, and a potential is applied during the readout, the electrode acts as a gate and dedopes/inhibits the neighbouring synapses (Fig. S9c), hence reducing the output.
\section{Conclusion}
We propose field-directed polymerisation (FDP) as a novel approach to grow polymeric dendritic connections composed of many artificial synapses. FDP allows to guide the polymerization, tune the branching, and control both resistance and capacitance. The growth resembles the synaptogenesis process and the synapses show plastic responses triggered by time-dependent electrical stimuli. We demonstrate key features needed in synaptic electronics and neuromorphic computing such as spike-timing dependent plasticity, long-term potentiation, and depression from the millisecond to week timescale, and employ FDP to achieve Pavlovian conditioning and Hebbian learning. We achieve this by controlling both the temporal pattern of excitations and material properties of the neural networks: specifically, we change the growth parameter to influence the resistance and the capacitance, and we use them to affect the neuromorphic functions. Finally, we combine these features to show a new device concept capable of recognizing numeric patterns.

\bibliography{refs}
\bibliographystyle{naturemag}

\section*{Methods}
Fibres of the conductive poly-3,4-ethylenedioxythiophene doped with hexafluorophosphate (PEDOT:PF$_6$) are prepared by field-directed polymerisation in an electrolyte solution under AC signals (10 to 200 Hz) with a HP 8114A pulse generator between two gold electrodes.
The electrolyte solution of acetonitrile contains , 1 mM TBAPF$_6$, and 50 mM EDOT. Note that the solvent dissolves the salt and the monomer, but poorly dissolves the resulting polymer.
The highly doped PEDOT:PF6 is formed with dark blue colour and rinsed in isopropanol, and annealed at 110 degree for 10 min.
he electrical characteristics are measured  with a pair of Keithley SMUs (models 2600 and 2400) using the software SweepMe! (sweep-me.net) with the device immersed in an aqueous solution containing 1 mM NaCl. To prevent parasitic currents, an insulating layer (SU8, microchem) is patterned on top of the substrate. The action potential-like wave form was programmed using an arbitrary function generator.
For the number recognition, it is important to note that the capacitance of the bare Au electrodes was not enough to provide a large gating effect: to overcome this limitation, all the electrodes were covered in PEDOT:PF$_6$ through DC electropolymerisation (3 V for 3 s) before the training epoch. 
For the decay characterisation, the devices are stored in an aqueous electrolytic solution of 1 mM NaCl over the whole time.

\section*{Data Availability}
The data that support the findings of this study are available from the corresponding author upon reasonable request.

\end{document}


\maketitle
\begin{affiliations}
 \item Dresden Integrated Center for Applied Physics and Photonic Materials (IAPP), Technische Universit{\"a}t Dresden, 01062 Dresden, Germany
\end{affiliations}

\linenumbers
\section*{Supplementary Information}

 \begin{figure}
 	\includegraphics[height=7cm]{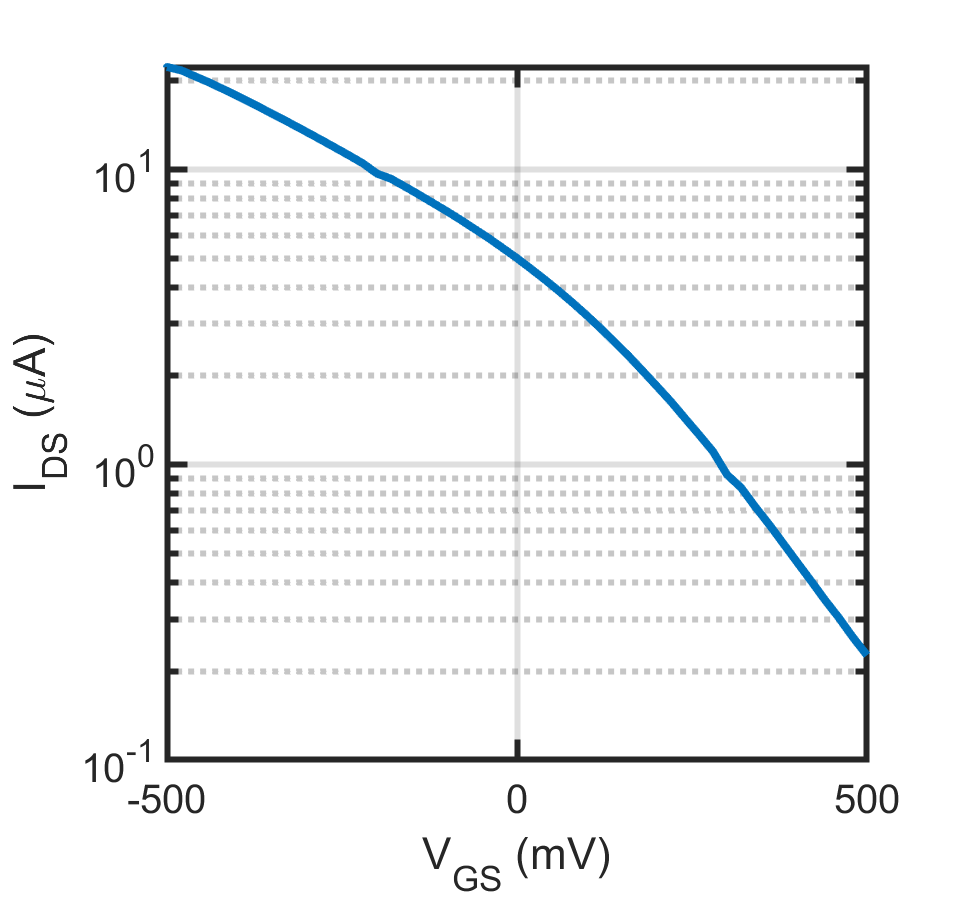}
 	\caption{\textbf{ | Characterisation of a PEDOT artficial synapse as an OECT:} Transfer curve relative to the device in Fig. 2d after the third update of the synaptic weight (V$_{DS}$=500 mV).}
 \end{figure}

\begin{figure}
	\includegraphics[width=\textwidth]{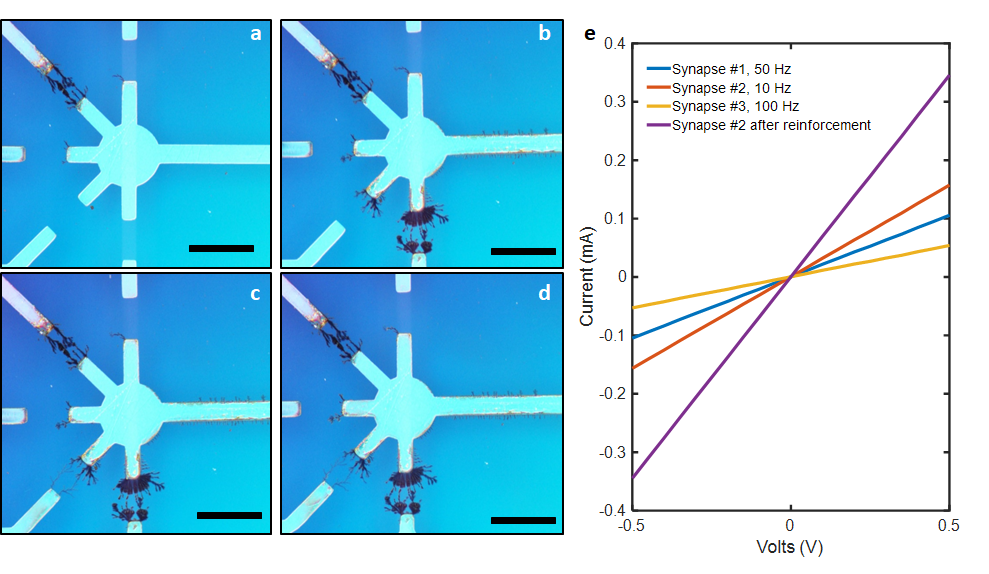}
	\caption{\textbf{ | Directed growth and strengthening}: a) a synapse is grown between a selected input electrode and a neuron-like electrode with an AC signal of 50 Hz. \textbf{b} A second and \textbf{c} a third electrodes are selected and synapses are grown at 50 Hz and 100 Hz respectively. \textbf{d} A synapse already exisiting can be reinforced: in this case, the one grown in panel b. e) IV curve of the synapses. Scale bar is 200 $\mu$m.}
\end{figure}

 \begin{figure}
 	\includegraphics[height=7cm]{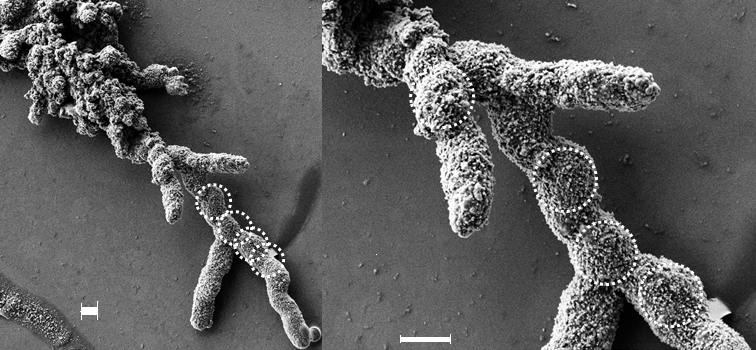}
	\caption{\textbf{| Fibre's morphology:} Scanning electron microscopy picture of a PEDOT:PF6 fiber. Nucleation sites are visible as spherical structures. Scale bar is 2$\mu$m. }
\end{figure}

\begin{figure}
	\includegraphics[height=7cm]{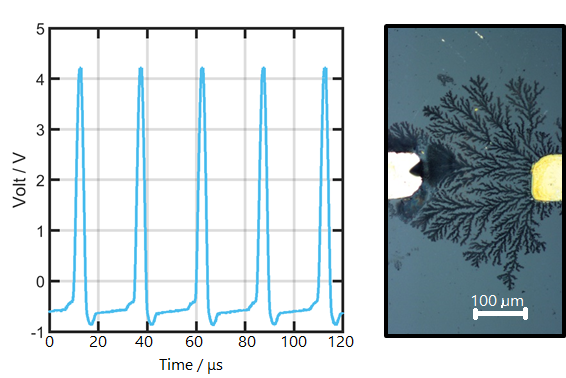}
	\caption{\textbf{ | Artificial synapse grown with an action potential-like waveform.} A signal resembling the action potential at (5 V, 50 Hz) was used to grow an artificial synapse with FDP.}
\end{figure}

 \begin{figure}
		\includegraphics[width=\textwidth]{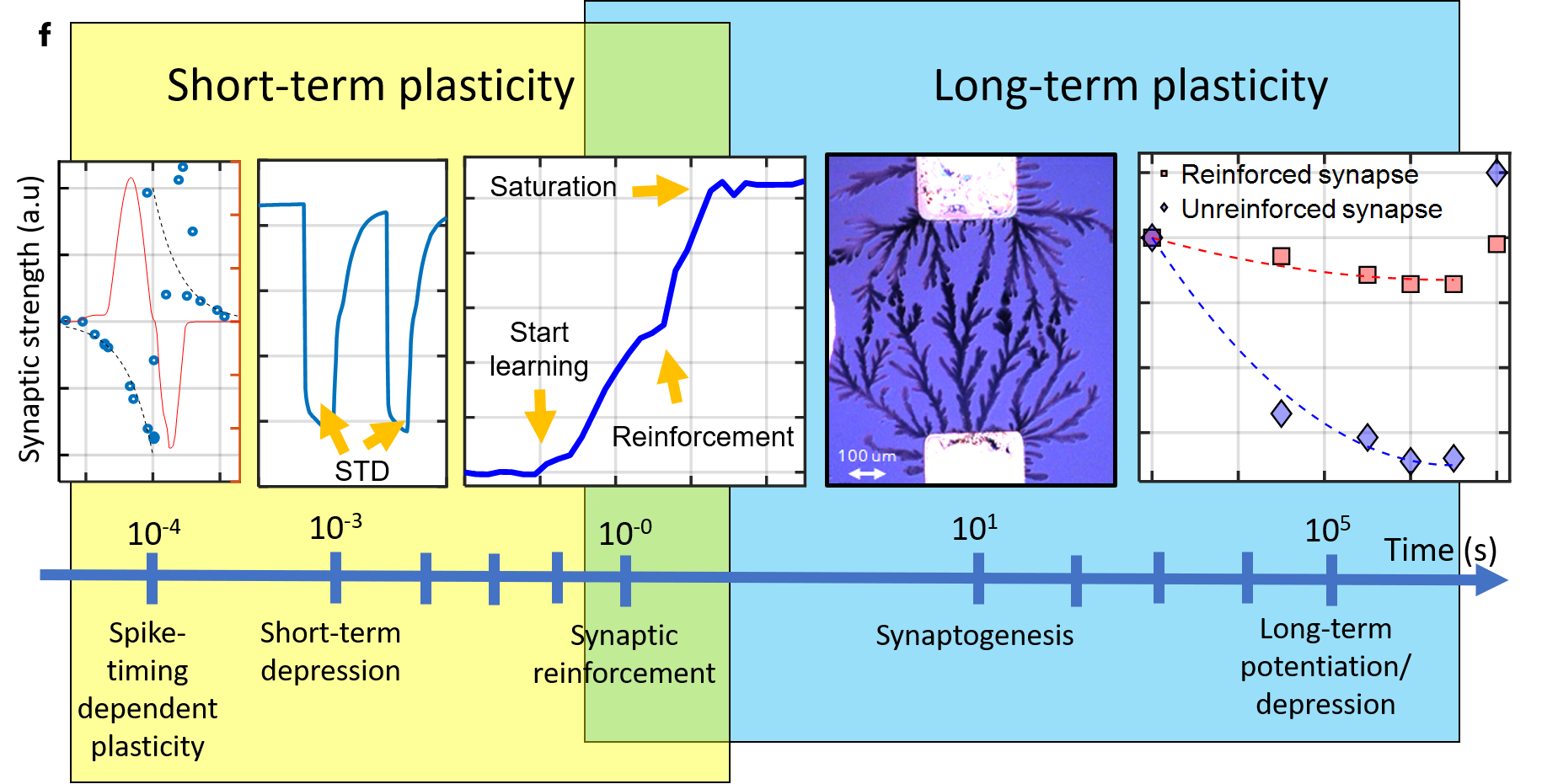}
	\caption{\textbf{ | Plasticity over different timescales.} We achieve plastic effects that span from short-term plasticity (ms range) up to days and week scale (long-term plasiticity)}
\end{figure}

\begin{figure}
		\includegraphics[height=7cm]{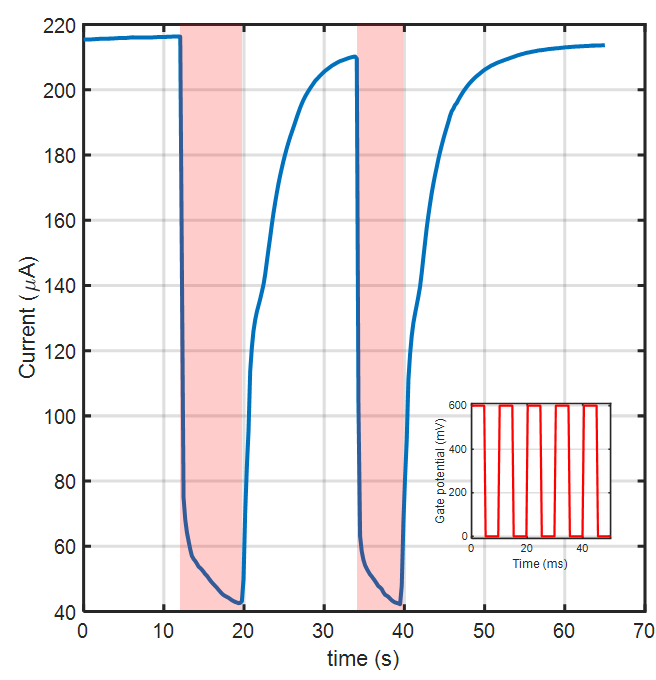}
		\caption{\textbf{ | Short-term depression}. By applying a positive pulsed signal (600 mV, 100 Hz, see inset) to a third electrode between 10 and 20 s, and between 35 and 40 s, the channel is dedoped and short-term plasticity at the millisecond scale is mimicked.}
\end{figure}

 \begin{figure}
	\includegraphics[width=\textwidth]{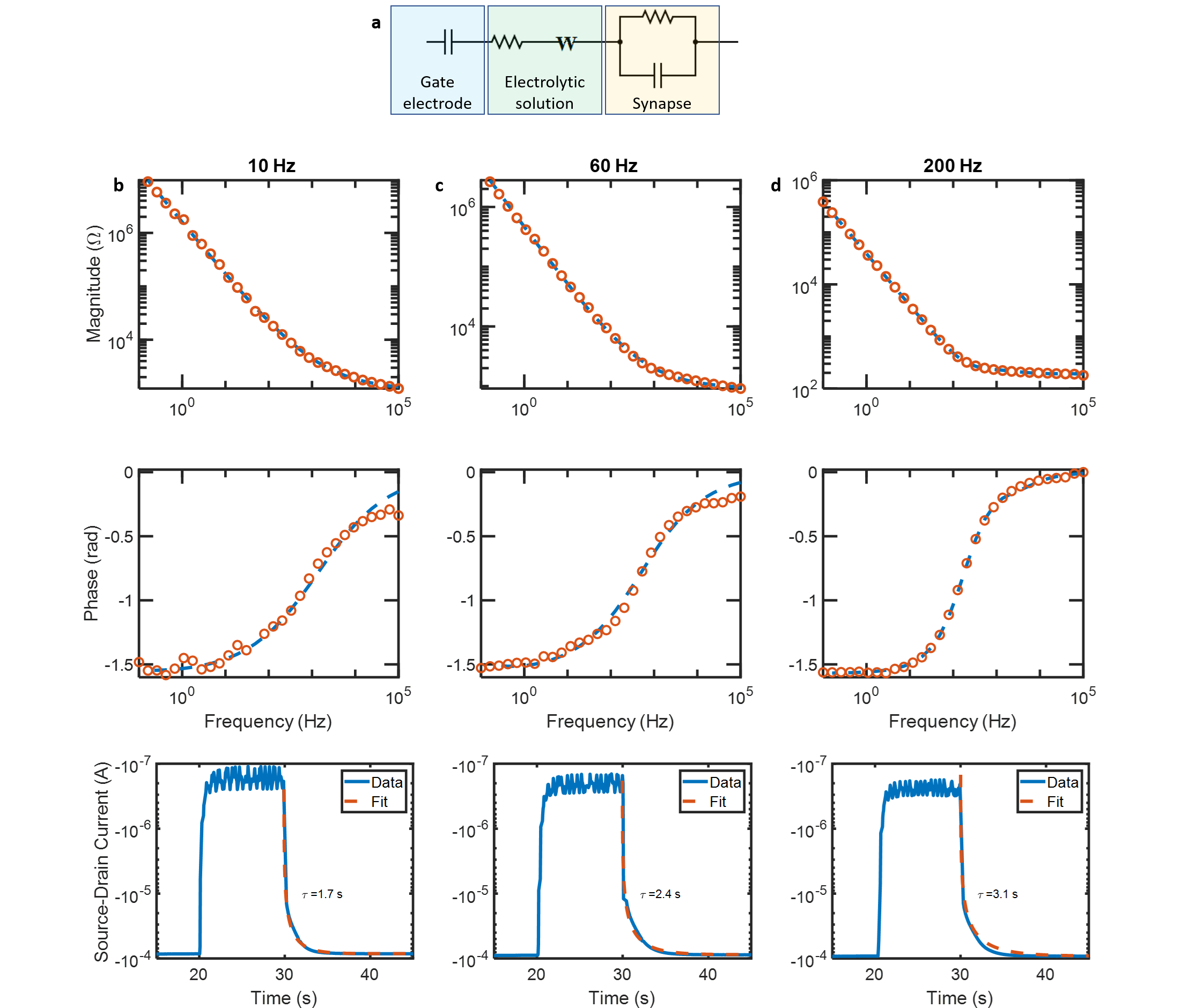}
	\caption{\textbf{ | Impedance spectra of synapses grown at different frequency} The different morphology that the synapses acquire when varying the polymerization frequency (See Figure 1c) is reflected in a different capacitance. \textbf{b} shows the circuit used to fit the data: the gate and electrolytic solution are kept fixed (within an error of 20\%). The top and middle panel of \textbf{b} show the data acquired with impedance spectroscopy and the relative fit. It follows that the devices have a large capacitance of 0.10, 0.39 and 1.93 $\mu$F for 10, 60 and 200 Hz respectively. This, in turn, affects the charging time ($\tau$=RC time), as shown in the bottom panel.} 	
\end{figure}
 \begin{figure}
	\includegraphics[width=\textwidth]{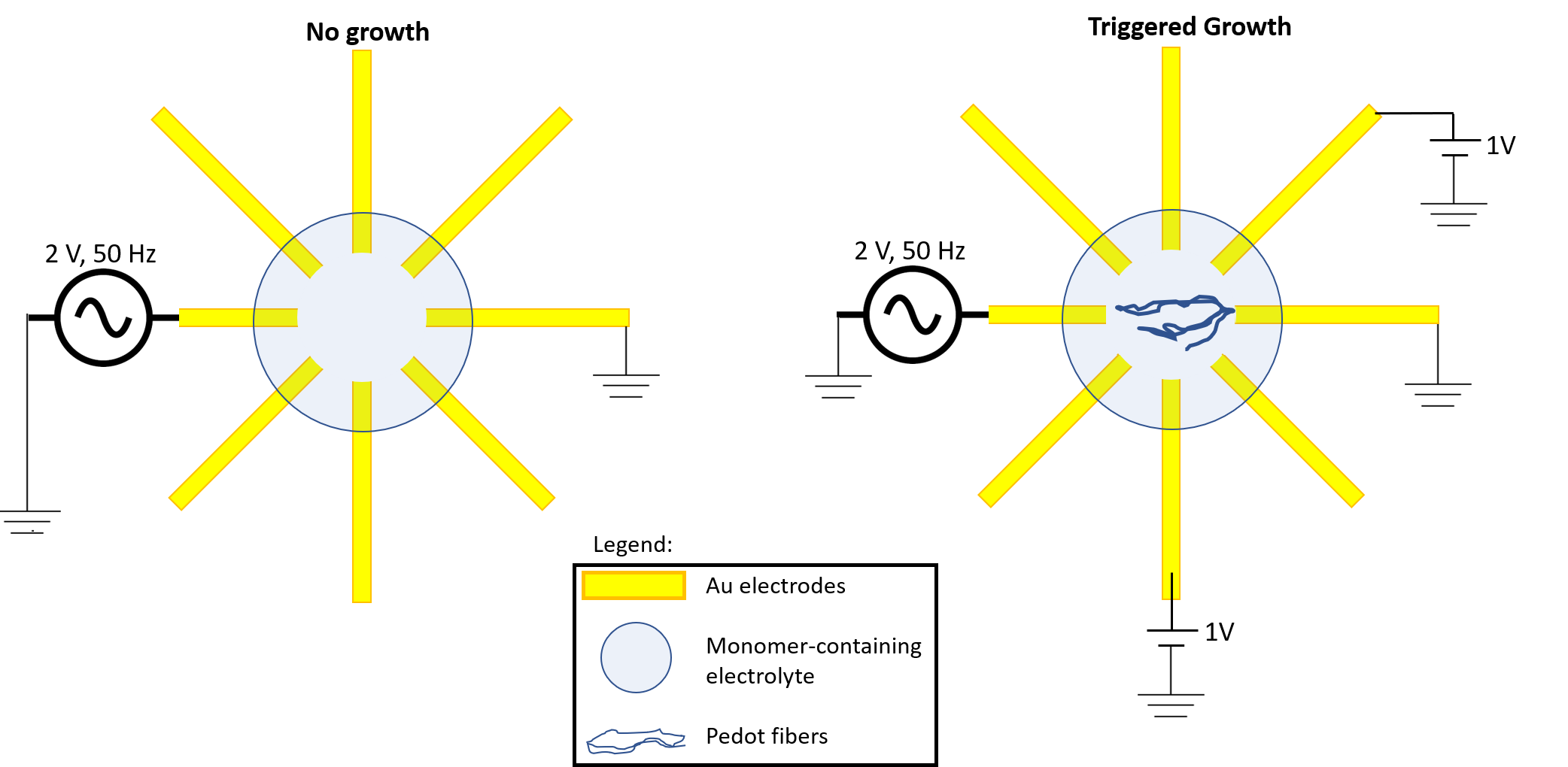}
	\caption{\textbf{ | Schematics of the mechanism of triggered growth}: on the left, an AC signal is applied but, being insufficient for the oxidation of the monomer, it does not cause the formation of the fiber. The growth can be triggered without changing its potential, rather, applying a voltage to neighbouring electrodes/neurons (right panlel). Hence, it is possible to grow at low voltage and control it by shaping the field in the electrolyte.}	
\end{figure}

\begin{figure}
	\includegraphics[width=\textwidth]{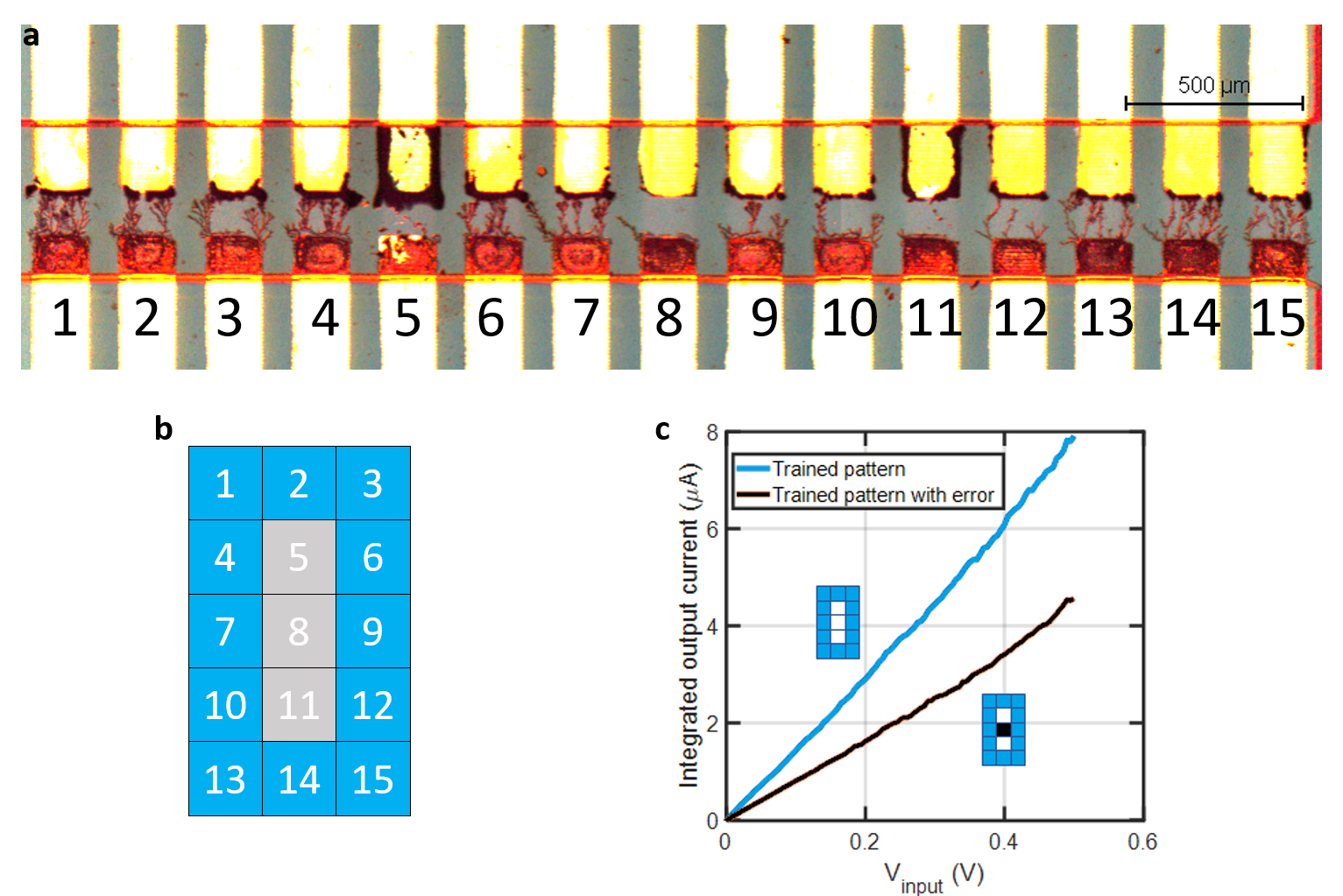}
	\caption{\textbf{| Pattern recognition system. a,} Pattern recognition device after the learning epoch for the number 0. \textbf{b}, Trained pattern corresponding to panel a. \textbf{c}, Output current for the exact pattern and for a single-pixel error pattern (number 8): the extra input does not carry current to the output. Rather, it attracts negatively charged dopants in the solution and dedopes neighbouring synapses, hence reducing the total output.}
\end{figure}
\section*{Data Availability}
The data that support the findings of this study are available from the corresponding
author upon reasonable request.